\documentclass[aps,prl,twocolumn,groupedaddress,showpacs]{revtex4}

\usepackage{epsf,graphicx}

\begin{document}


\title{Generation of stable ultra-relativistic attosecond electron bunches via the laser wakefield acceleration mechanism}



\author{M.J.H. Luttikhof}
\email{m.j.h.luttikhof@utwente.nl}
\author{A.G. Khachatryan}
\author{F.A. van Goor}
\author{K.-J. Boller}

\affiliation{Faculty of Science and Technology and MESA+ Institute, University of Twente, P.O. Box 217, 7500 AE Enschede, The Netherlands}


\date{\today}

\hyphenation{Euroleap}

\begin{abstract}
In recent experiments ultra-relativistic femtosecond electron bunches were generated by a Laser Wakefield Accelerator (LWFA) in different regimes. Here we predict that even attosecond bunches can be generated by an LWFA due to the fast betatron phase mixing within a femtosecond electron bunch. The attosecond bunches are stable outside the LWFA and can propagate in vacuum many tens of centimeter without significant change in their duration. Our calculations show that evidence for the formation of attosecond bunches can be found in the spectrum of coherent betatron radiation from LWFA's.
\end{abstract}



\maketitle

The generation of attosecond electron bunches would be of great interest to provide a new and unique tool for modern physical research. The potential of such bunches ranges from electron microscopy with attosecond resolution, to the generation of attosecond X-ray beams, for investigating physical, chemical and biological processes on the attosecond timescale.

The shortest bunches to date are available from Laser Wakefield Accelerators (LWFA's) \cite{tajima}. In recent experiments exploiting different parameter regimes \cite{mangles, leemans, faure2, geddes2, osterhoff, hafz, geddes, faure}, ultra-relativistic femtosecond (fs) electron bunches were generated by an LWFA. Here we predict that even attosecond bunches can be generated by an LWFA due to the fast betatron phase mixing withing a femtosecond electron bunch. The attosecond bunches are stable outside the LWFA, which means that they can propagate through vacuum over distances of many tens of centimeters, without a significant change of duration. We also predict how such bunches can be identified in the optical spectrum emitted during the acceleration process.

Several schemes for the generation of attosecond electron bunches have been proposed so far. Examples are an inverse free-electron-laser process \cite{sears}, the interaction of high intensity laser pulses with overdense plasma \cite{naumova}, the acceleration of electrons with a short tailored laser pulse \cite{stupakov}, the slicing of an electron bunch with a laser pulse \cite{dodin}, the interaction of an ultra-short laser pulse with a nanofilm \cite{kulagin}, with a wire or a plasma slice \cite{ma}, or the interaction of a PW radially-polarized laser pulse with a sub-micron droplet of a high-Z material \cite{karmakar}. However, all these schemes are of limited attractivity, because they either require large accelerator structures \cite{sears} or high intensities in the order of $10^{20}-10^{22}$ W cm$^{-2}$ (normalized amplitude, $a_0$, 10-100) \cite{naumova, stupakov, dodin, kulagin, ma, karmakar}, such as available only from rather exclusive (Petawatt) laser systems. A strong disadvantage is that these schemes would deliver bunches of rather low energy (a few to a few tens of MeV's) which would make it difficult to keep the duration of such bunches and apply them due to space charge effects. In some schemes the bunches would suffer from a limited life time (about 10 fs \cite{kulagin}) or a limitation of charge to well below a pC \cite{sears, stupakov, karmakar}.

Our scheme of attosecond bunch generation, presented in this Letter, has the advantage that it is based on the nowadays well-known technique of LWFA, and that the attosecond bunches are accelerated to GeV energies with the use of a Gaussian laser pulse with $a_0 \sim 1$ from a commercially available TW laser system. We show that the generated attosecond bunches also live long when they have left the laser wakefield.

The working principle of our scheme is based on a frequency sweep (chirp) of the betatron oscillation of the local bunch radius along an fs electron bunch during laser wakefield acceleration. Betatron oscillations \cite{khachatryan9} are initiated by a mismatch of the electron bunch radius with regard to the focusing field, and the local bunch radius will oscillate between some minimum and maximum value determined by the bunch emittance and the focusing gradient \cite{humphries}. The oscillation frequency of the radius is twice the betatron frequency given by $\omega_\beta = \omega_p \sqrt{{f}/{\gamma}}$, where $\omega_p$, $f$ and $\gamma$ are, respectively, the plasma frequency, the focusing gradient and the relativistic factor \cite{khachatryan9}. It can be expected that such oscillations occur because the bunch becomes naturally mismatched, due to the longitudinal change in the focusing gradient and due to the increase of the relativistic factor during acceleration \cite{khachatryan9}. Our investigations show that, despite the extremely short bunch duration in an LWFA, the different parts of the bunch with different longitudinal positions in the direction of propagation undergo betatron oscillations at different frequencies, mainly due to the longitudinal variation of the focusing gradient, but also due to the longitudinal variation of the relativistic factor. This dynamics leads to modulation of the radius of the fs bunch on an attosecond scale. This way, sharp peaks of attosecond duration are formed in the electron density, during the acceleration of electrons to ultra-relativistic energies.

Our calculations over a wide range of experimental parameters, such as typical for channel-guided LWFA's, clearly indicate that the described attosecond dynamics is a very general and intrinsic feature of LWFA's. Moreover, the signature of attosecond bunches may even be present in the radiation spectra of currently performed experiments, as is discussed below.

In order to demonstrate more details of the described bunching in an LWFA, we model the dynamics with an fs electron bunch that has been generated via some injection mechanism (for example, in the bubble regime \cite{leemans} or using a plasma density gradient injection \cite{geddes2}) and where further acceleration occurs in a channel-guided LWFA. For the latter, one could tink of a two-stage LWFA, where the fs bunch is formed in a gas jet and further accelerated in a plasma channel \cite{khachatryan9}. To describe the dynamics with typical parameters, we assume that the bunch has initially an energy of 51 MeV ($\gamma = 100$), an energy spread of 3\%, a full-width-at-half-maximum (FWHM) duration of 7.5 fs, a root-mean-square (rms) width of 1.3 $\mu$m in both transverse ($x$ and $y$) directions, and a normalized transverse emittance of 1.3 $\mu$m. The bunch is injected in a 48 mm long capillary discharge plasma channel with a radius $r_{\rm ch}$ of 61 $\mu$m and an on-axis electron concentration of $7 \times 10^{17}$ cm$^{-3}$ corresponding to a plasma wavelength $\lambda_p$ of 40 $\mu$m. The radial unperturbed electron density profile $n_p(r)$ has the typical parabolic profile \cite{esarey3}. The Gaussian laser pulse that drives the wakefield in this plasma channel has a FWHM duration of 35 fs, is focused to a waist of 38 $\mu$m, and has a peak intensity of $1.7 \times 10^{18}$ W$\,$cm$^{-2}$ ($a_0 = 0.9$). The laser pulse is mismatched to the plasma channel to suppress the oscillations in its intensity during propagation in the channel due to the self-focusing, as was done in \cite{luttikhof2}. The laser pulse is linearly polarized and has a central wavelength of 0.8 $\mu$m. The calculations were carried out using the fully relativistic particle code {\sc wake} \cite{antonsen}, which includes also the laser pulse dynamics.

\begin{figure}
\includegraphics[width=0.5\textwidth]{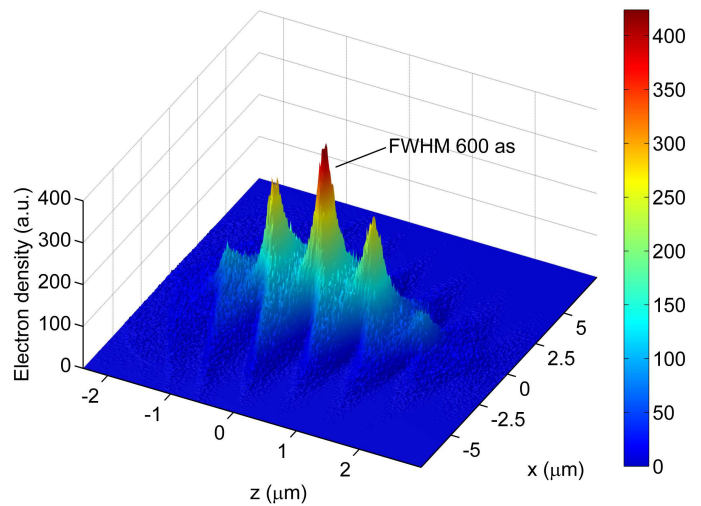} 
\caption{\label{fig:2ndstage} Electron density distribution of the accelerated fs electron bunch (in arbitrary units) at the exit of a laser wakefield accelerator.}
\end{figure}

The electron bunch, which initially has a Gaussian density distribution, is injected in the first maximum of the accelerating field behind the laser pulse. In Fig. \ref{fig:2ndstage} the bunch is shown, when it has just left the 48 mm long plasma channel, where it has been accelerated to 655 MeV with an energy spread of 10\%. 

As can be seen, the fs electron bunch with initially a Gaussian distribution is transformed to a few sub-bunches of attosecond duration. The duration of the sub-bunches is approximately 600 as. This structure is formed due to a transverse change in the distribution of electrons, caused by the betatron oscillations of the local bunch radius.

An attosecond structure can be formed in other ways, for example, via an external injection scheme like injection in front of the laser pulse \cite{khachatryan} or injection at an angle \cite{luttikhof2}. Here we will give an example of the attosecond structure formed for the scheme where the bunch is injected in front of the laser pulse. In this calculation a 250 fs (FWHM) long electron bunch, which is accelerated to a kinetic energy of 2.9 MeV with a standard radio-frequency photo-cathode linac, is injected into a plasma channel. The bunch has a Gaussian distribution in both longitudinal and transversal directions and a 0.7\% energy spread. We assume that the bunch charge is below the beam-loading limit, so that the space charge effects can be neglected. The electron bunch is focused into the plasma channel to an rms radius of 38 $\mu$m. The normalized emittance of the injected bunch is 0.6 $\mu$m in both transverse coordinates $x$ and $y$. The plasma channel and the laser pulse parameters are the same as in the first example. Immediately after the electrons enter into the plasma channel, the high intensity laser pulse follows. In the plasma, the electrons will be overtaken by the laser pulse, because the velocity of the bunch is smaller than that of the pulse. In this process, a considerable part of the bunch will be trapped, compressed and accelerated in the first accelerating region behind the laser pulse \cite{khachatryan}.

\begin{figure}
\includegraphics[width=0.5\textwidth]{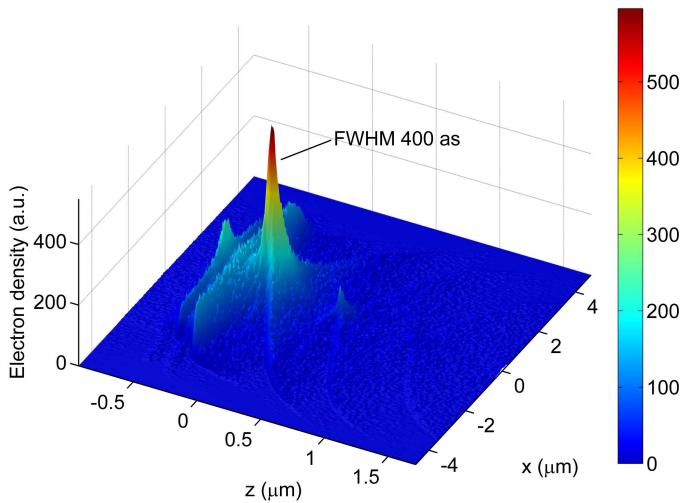}
\caption{\label{fig:infront} Electron density distribution of the accelerated fs electron bunch (in arbitrary units) where the initial bunch is externally injected in front of the laser pulse.}
\end{figure}

The typical structure of the trapped bunch that is formed and propagates in the wakefield can be seen in Fig. \ref{fig:infront}. Here we see the electron bunch as it comes out of the 48 mm long plasma channel. In this distance the electrons have been accelerated to an energy of 690 MeV with an energy spread of 5.5\%. The rms radius of the accelerated fs bunch is 1.4 $\mu$m and its transverse emittance is 2.4 $\mu$m. Again, a finestructure, caused by the betatron phase mixing, is formed inside the bunch. Due to the laser pulse dynamics, which especially plays a big role during the trapping of the bunch, only one strong attosecond density peak shows up. The total FWHM bunch duration is 5.4 fs. The FWHM duration of the high density sub-bunch is approximately 400 as.

As was mentioned in the introduction, the stability of the attosecond bunches in the vacuum, after leaving the LWFA, is crucial. Now we will show that the generated attosecond structure will be preserved in the vacuum even after propagation of tens of centimeters. To explain this, consider a bunch slice with a fixed longitudinal position, $z$, relative to the bunch center. The radius of the slice (the local bunch radius), $\sigma$, evolves in the vacuum according to \cite{khachatryan9}
\begin{equation} \label{eq:evolutionofradius}
\sigma = \sigma_* \sqrt{1 + \frac{(z_{\rm pr} - z_*)^2}{Z_b^2}},
\end{equation}
where $Z_b = \gamma \sigma_*^2/\varepsilon_n$ is the characteristic distance on which the radius grows or the ''Rayleigh'' length and $z_{\rm pr}$ is the propagation distance. The Rayleigh length depends on the spot size in the focus or the waist $\sigma_*$, the relativistic factor $\gamma$ and the normalized emittance $\varepsilon_n$. The waist of the bunch slice is given by $\sigma_*^2 = \varepsilon_n^2/\gamma^2 h$, where $h = \varepsilon_n^2/\gamma^2 \sigma_p^2 + \sigma_p'^2$ with $\sigma_p$ and $\sigma_p' = d\sigma_p/dz$ the bunch radius and the divergence at the exit of the plasma. When the bunch leaves the wakefield it can still have focusing properties, which means that it will focus at a distance behind the channel given by $z_* = \sigma_p \sigma_p'/h$. The local radius in the structured bunch coming from the LWFA can be approximated by \cite{khachatryan9}
\begin{equation} \label{eq:structure}
\sigma_p^2 = \sigma_0^2 + \sigma_1^2 \sin{\left(\frac{2 \pi z}{\lambda_b} + C\right)},
\end{equation}
where  $\sigma_0$, $\sigma_1$ and $C$ are constants and $\lambda_b$ is the distance between neighbor sub-bunches. With the local bunch radius and its divergence at the exit of plasma given by expression (\ref{eq:structure}), equation (\ref{eq:evolutionofradius}) can be applied to predict the radius of each slice in vacuum. Simple analysis reveals that the Rayleigh length, $Z_b$, changes monotonically throughout the fs bunch having minimum (maximum) for minimum (maximum) $\sigma_p$ (in these cases $\sigma_p' = 0$). This means that the part of the bunch with smaller radius will diverge stronger. From this we can conclude that in the vacuum there will be a transition where the structure is reversed, which means that the parts with largest radius become the parts with smallest radius and vice versa. Another way to look at this, is by looking at the transverse momenta of the electrons, which has a maximum for the parts with minimum radius and vice versa. The reversing process, confirmed by our simulations, typically happens in the first few millimeters behind the plasma channel. After this process is finished, the attosecond bunch structure stays stable during the propagation over many tens of centimeters in vacuum.

\begin{figure}
\includegraphics[width=0.5\textwidth]{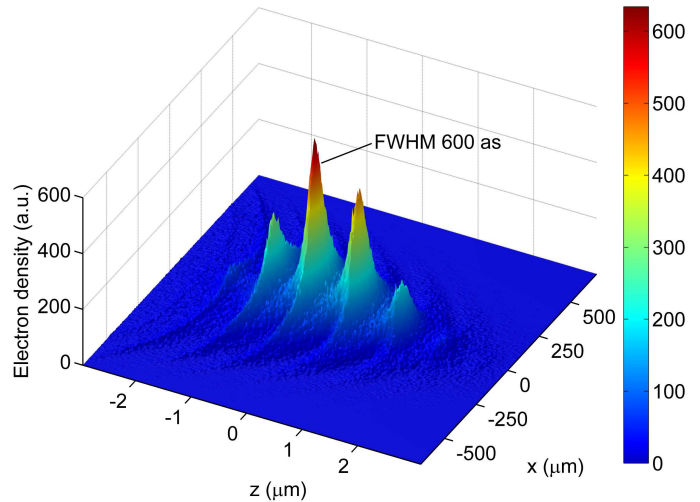}
\caption{\label{fig:2ndstage10cm} Electron density distribution of the accelerated fs electron bunch (in arbitrary units), presented in Fig. \ref{fig:2ndstage}, after propagating 10 cm in vacuum.}
\end{figure}

The propagation in vacuum, of the bunch depicted in Fig. \ref{fig:2ndstage}, was calculated using the {\sc gpt} code \cite{gpt}. We found that for a typical bunch charge of a few tens of pC, space charge effects practically do not play a role and can be neglected. We calculated what the bunch profile will be after 10 cm of propagation in the vacuum. The resulting bunch density is plotted in Fig. \ref{fig:2ndstage10cm}. The rms bunch radius has grown to 145 $\mu$m, but as discussed above, the attosecond structure is maintained, with the difference that the maxima and minima of the electron density are now reversed.

\begin{figure}
\includegraphics[width=0.5\textwidth]{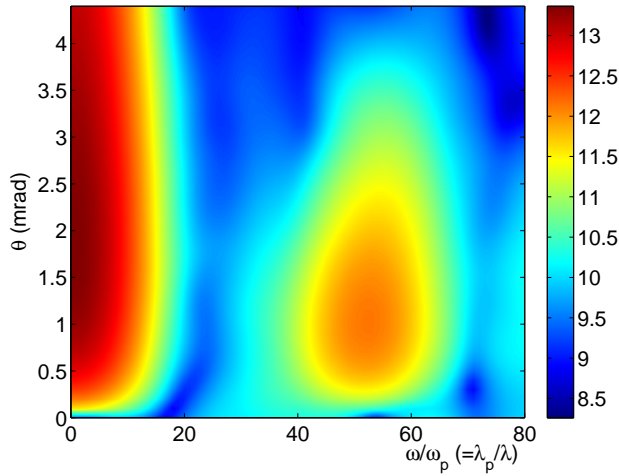}
\caption{\label{fig:radiation} Coherent betatron radiation emitted by an electron bunch in a laser wakefield accelerator. The spectral and angular distribution of the normalized radiation energy is given on a logarithmic scale.}
\end{figure}

The formation of attosecond structure is an intrinsic effect of an LWFA and is something that might even occur in currently running experiments. However measuring the density distribution of an LWFA bunch is something that can not be easily done. Here we will show that attosecond structure in a bunch in an LWFA can be identified by looking at the emitted coherent betatron radiation \cite{khachatryan8}. We have calculated the radiation for the case presented in Fig. \ref{fig:2ndstage} the same way as we have done in \cite{khachatryan8}. In Fig. \ref{fig:radiation} we have depicted the angular and spectral distribution of the radiated energy normalized to $e^2/4 \pi^2 c$ (here $e$ is the elementary charge and $c$ is the speed of light), in the logarithmic scale; $10^4$ particles were used for this simulation. Note that the radiation at frequencies lower than $\omega_p$ will be absorbed in the plasma and cannot be observed outside the plasma channel. The radiation is confined to very small observation angles in the order of $1/\gamma_m$ \cite{khachatryan8}, where $\gamma_m$ is the average relativistic factor of the electrons at the exit of the accelerator. The strong radiation at low frequencies seen in Fig. \ref{fig:radiation} is the coherent betatron radiation from the fs electron bunch, which we studied in \cite{khachatryan8}. This radiation scales approximately as $\exp{[-\left(\omega \sigma_z/c\right)^2]}$ (here $\omega$ is the frequency and $\sigma_z$ the rms bunch length) that agrees with the form-factor formalism \cite{nodvick}, according to which the energy radiated by a bunch is proportional to $N_e \left[1 + \left(N_e-1\right)f\right]$, where $N_e$ is the number of particles in the bunch and $f$ is the form-factor. There is also weaker emission that is peaked at higher frequencies, which we found from the simulations to be attributed to the fine attosecond structure within the fs bunch. To verify this, we have calculated analytically the form-factor of a structured fs bunch, modelling it as a Gaussian bunch with a local radius that satisfies expression (\ref{eq:structure}). Then, we found that the emitted radiation has a second peak at the frequency
\begin{equation} \label{eq:radiationpeak}
\omega_* \approx \frac{c \pi}{\lambda_b} \left(1 + \sqrt{1 + \frac{2 \lambda_b^2}{\pi^2 \sigma_z^2}}\right).
\end{equation}
For the electron bunch depicted in Fig. \ref{fig:2ndstage}, according to (\ref{eq:radiationpeak}), there should be a peak in the spectrum of the betatron radiation at a wavelength of 0.77 $\mu$m,  which corresponds to $\omega_* \approx 52 \omega_p$. This agrees very well with the value predicted by the numerical calculation (see Fig. \ref{fig:radiation}).

In conclusion, our calculations predict that a laser wakefield accelerated electron bunch can acquire a finestructure at the attosecond scale. It turns out that this is an intrinsic property of the channel-guided LWFA. We showed that structure is formed in a second-stage LWFA and in the case of external bunch injection in front of the laser pulse. From other simulations we performed, it became clear that an attosecond structure is also formed for external injection at an angle \cite{luttikhof2}. When the bunch leaves the wakefield and enters the vacuum, the attosecond structure is reversed in a few millimeters of propagation. The bunch diverges while further propagating in the vacuum, but the attosecond structure remains present and is stable during a long propagation distance.

\begin{acknowledgments}
This work was supported by the Dutch Ministry of Education, Culture and Science (OC\&W) and the European Community-New and Emerging Science and Technology Activity (project EuroLEAP, contract number 028514). We acknowledge professor P. Mora for providing the {\sc wake} code and detailed explanations of it.
\end{acknowledgments}


\end{document}